\documentclass[journal]{IEEEtran}

\ifCLASSINFOpdf
\else
\fi

\usepackage{graphicx}
\usepackage{makecell}
\usepackage{amsmath}
\usepackage{amsfonts}
\usepackage{multirow}
\usepackage{array}
\usepackage{enumerate}
\usepackage{stfloats}
\usepackage{cite}
\usepackage{booktabs}
\usepackage{bm}
\begin{document}
%
\title{Low-Latency and Low-Complexity MLSE for Short-Reach Optical Interconnects}
%
%
%

\author{Mengqi Guo, Ji Zhou, Haide Wang, Changyuan Yu, Xiangjun Xin, and Liangchuan Li
\thanks{This work was supported in part by National Key R\&D Program of China under Grant 2023YFB2905700, in part by Science Research Project of Hebei Education Department under Grant QN2025007, in part by National Natural Science Foundation of China under Grants 62371207 and 62005102, and in part by Young Elite Scientists Sponsorship Program by CAST under Grant 2023QNRC001. \emph{(Corresponding author: Ji Zhou.)}}
\thanks{Mengqi Guo is with School of Information Science and Technology, Shijiazhuang Tiedao University, Shijiazhuang 050043, China.}
\thanks{Ji Zhou, Xiangjun Xin, and Liangchuan Li are with Aerospace and Informatics Domain, Beijing Institute of Technology, Zhuhai 519088, China.}
\thanks{Haide Wang is with School of Cyber Security, Guangdong Polytechnic Normal University, Guangzhou 510665, China.}
\thanks{Changyuan Yu is with Department of Electrical and Electronic Engineering, The Hong Kong Polytechnic University, Hong Kong SAR, China.}

}

\maketitle

\begin{abstract}
To meet the high-speed, low-latency, and low-complexity demand for optical interconnects, simplified  maximum likelihood sequence estimation (MLSE) is proposed in this paper. Simplified  MLSE combines computational simplification and reduced state in  MLSE.  MLSE with a parallel sliding block architecture reduces latency from linear order to logarithmic order. Computational simplification reduces the number of multipliers from exponential order to linear order. Incorporating the reduced state with computational simplification further decreases the number of adders and comparators. The simplified  MLSE is evaluated in a 112-Gbit/s PAM4 transmission over 2-km standard single-mode fiber. Experimental results show that the simplified  MLSE significantly outperforms the FFE-only case in bit error ratio (BER) performance. Compared with simplified 1-step MLSE, the latency of simplified  MLSE is reduced from 34 delay units in linear order to 7 delay units in logarithmic order. The simplified scheme in  MLSE reduces the number of variable multipliers from 512 in exponential order to 33 in linear order without BER performance deterioration, while reducing the number of adders and comparators to 37.2\% and 8.4\%, respectively, with nearly identical BER performance.
\end{abstract}


%
\IEEEpeerreviewmaketitle

\section{Introduction} \label{section1}
\IEEEPARstart{D}{riven} by the data-intensive applications such as artificial intelligence, cloud computing, and Internet of Things, the demand for data centers with high-speed, low-complexity, and low-latency grows sharply \cite{intc1,intc2,intc3}. Intensity modulation and direct-detection (IM/DD) has the advantage of low cost and simple structure, which has been widely applied in the high-speed and short-reach optical interconnects for data center \cite{IMDD1,IMDD2,IMDD3}. Low-cost and small-footprint electrical/optical devices are favored in short-reach optical interconnects, so that the high-speed signals inevitably suffer from high-frequency distortion caused by bandwidth-limited devices. In order to compensate for the inter-symbol interference (ISI) caused by bandwidth-limited devices, the direct detection faster than Nyquist (DD-FTN) scheme based on a feed-forward equalizer (FFE), a noise whitening post-filter, and maximum likelihood sequence estimation (MLSE) is proposed \cite{zhong}. In this work, 4$\times$112Gb/s PAM4 transmission over 2-km standard single-mode fiber (SSMF) is achieved by utilizing MLSE for performance improvement. Owing to the significant ISI elimination performance, the DD-FTN scheme and similar schemes using polynomial nonlinear equalizer or decision-feedback equalizer have been widely applied in optical interconnects \cite{guo,MLSE1,MLSE2,MLSE3,MLSE4,MLSE5}.

In the DD-FTN scheme, MLSE not only eliminates the residual ISI from FFE, but also eliminates the ISI introduced by the post-filter. However, the performance improvement achieved by MLSE comes with high computational complexity of the Viterbi algorithm. The research about reducing the computational complexity of MLSE is mainly divided into two categories: reducing the number of states and reducing the number of multipliers. To reduce the number of states, the output of FFE can be used as a pre-decision value to reduce the number of candidate states in MLSE \cite{hoon,jing,Hiroki}. Through the pre-decision value, the candidate states with high probability are reserved, while the other states are discarded to reduce the trellis size. Moreover, the decision region whether the reduced-state MLSE is applied can be iteratively determined by error rate \cite{li}. To reduce the number of multipliers, in the branch metric (BM) calculation of MLSE, the lookup table is used to replace the convolution calculation \cite{dai}. The combination of lookup table and reduced state can further reduce the computational complexity \cite{jing2,li2}. The piecewise linear formula or absolute value can also be used to replace the squaring operation in BM calculation \cite{dai,cai}.

Besides the complexity, during the add-compare-select (ACS) calculation of MLSE, the current BM should continuously add the previous accumulated metrics in series. This serial calculation structure leads to the continuous increase of latency. With the rapid development of latency-constrained services, such as augmented reality, virtual reality, autonomous driving, and real-time gaming, the demand for low-latency data center optical interconnects is becoming urgent \cite{latency1,latency2,latency3}. Fettweis and Meyr first proposed the look-ahead M-step technique, which can effectively break the latency bottleneck of ACS operation in the conventional Viterbi algorithm \cite{M1,M2,M3}. Inspired by the look-ahead M-step technique, the layered look-ahead Viterbi decoding architecture is used to reduce the latency \cite{M4,M5}. Although the look-ahead M-step technique efficiently decreases the latency, its computational complexity still remains difficult to accept, especially when the modulation order of the signal, the number of channel coefficients in the post-filter, and the number of steps in the M-step are increased. 

To meet the high-speed, low-complexity, and low-latency demand for optical interconnects, the simplified  MLSE is proposed in this paper. The latency advantage of the  structure and the complexity reduction of simplified  MLSE are analyzed in detail. The bit error ratio (BER) performance of simplified  MLSE is experimentally verified in a 112-Gb/s PAM4 transmission over 2-km SSMF. The main contributions of this paper are as follows:
\begin{itemize}
\item The proposed simplified  MLSE combines the computational simplification and the reduced state in  MLSE. The latency is reduced from linear to logarithmic order through  MLSE with a parallel sliding block architecture. Computational simplification reduces the number of multipliers from exponential to linear order. Incorporating the reduced state further decreases the number of adders and comparators.
\item Experimental results show that compared with simplified 1-step MLSE (1S-MLSE), the latency of simplified  MLSE is decreased from 34 delay units in linear order to 7 delay units in logarithmic order without performance deterioration. Compared with  MLSE, the simplified  MLSE reduces the number of variable multipliers from 512 in exponential order to 33 in linear order, while the number of adders and comparators is reduced to 37.2\% and 8.4\%, respectively.
\end{itemize}

The rest of this paper is organized as follows. The principle of parallel 1S-MLSE is introduced in Section \ref{1-step}. The parallel  MLSE with low-latency property is described in Section \ref{2-step}. In Section \ref{2-step low complexity}, the proposed simplified  MLSE is described. In Section \ref{setup}, we present the experimental setup of 112-Gb/s PAM4 transmission over 2-km SSMF. In Section \ref{result}, we analyze the parameter settings, BER performance, latency and computational complexity of simplified  MLSE in the experiment. Finally, the paper is concluded in Section \ref{conclusion}.

\section{Parallel 1S-MLSE} \label{1-step}
\subsection{Signal Processing Procedure in MLSE}

MLSE employs the Viterbi algorithm to search for the most-likely state transition sequence in the state trellis \cite{viterbi1,viterbi2}, which is usually placed after the FFE and post-filter \cite{zhong}. The FFE can compensate for ISI at the cost of noise enhancement. The enhanced noise and the equalized signal are suppressed by the post-filter. Meanwhile, the post-filter can approximate the original channel response with the known shortened coefficients. Then, the signal filtered by the post-filter is fed into MLSE with the known ISI coefficients. The signal processing procedure of MLSE includes three steps. The three steps are BM calculation, ACS, and survivor path selection.

\textit{1) BM Calculation:} In the BM calculation, $s_n$ and $y_n$ are the transmitted signal and received signal at time index $n$. The vector $\boldsymbol{\alpha}$ represents the coefficients of the channel response, which includes the elements from $\alpha_0$ to $\alpha_{L-1}$, where $\alpha_0$ is equal to 1. For the PAM-$M$ signal, each transmitted signal $x_n$ at time index $n$ has $M$ possible different states. With $L$ coefficients of the channel response, the trellis has $M^{L-1}$ different states at each time index, and each previous state at time index $n-1$ extends $M$ different branches to the current state at time index $n$. Thus, the total number of BMs from time index $n-1$ to time index $n$ is $M^L$. To reduce computational complexity, a 2-tap post-filter ($L=2$) is often used, as it minimizes the number of BMs.

\textit{2) ACS Operation:} The objective of ACS is to find the best path at each state. Firstly, the accumulated value of previous BMs, which is named as path metric (PM) for each state, is added with the current BM extended from the corresponding state. Secondly, four different paths converge to the same state, and the path with the minimal PM is chosen at each state (an example is illustrated as a solid line). The minimal PM value at each state serves as the previous accumulated value for the next ACS calculation. Each ACS unit includes 4 adders and 3 comparators. In this 4-to-1 selection, the first two comparators are used for two 2-to-1 comparisons. Their results are input into the third comparator to obtain the 4-to-1 result. The signal $s_n$ includes 4 states, thus the total 4 ACS units at time index $n$ include 16 adders and 12 comparators.

\textit{3) Survivor Path Selection:} When the last 4 ACS calculations are finished, 4 different paths are reserved. Therefore, in the survivor path selection, the path with the minimal PM is chosen from the last 4 paths. This 4-to-1 selection still requires 3 comparators. Finally, the output of MLSE is determined based on the path with the minimal PM.

The received symbols calculated in the successive ACS should have a finite length. In the conventional MLSE, when the receiver receives a frame of symbols, the received symbols are processed by serially segmented sliding blocks. Each overlapped sliding block of $N$ symbols includes $R$ received symbols followed by $O$ overlapped symbols \cite{TB}. In each sliding block, when the ACS calculation completes at the last overlapped symbol, the final survivor path with $N=R+O$ symbols is obtained. Then, the $R$ received symbols are decoded into $D$ data symbols. Only the results of data symbols are reserved, and the overlapped symbols are removed. The PM of $D$ data symbols is transmitted to the next sliding block as the initial accumulated BM. 

\subsection{Parallel 1S-MLSE: Analysis of Latency and Complexity}
In practical MLSE implementation, the received symbols should be processed in parallel \cite{sliding}. The decision of the state at time index $n$ only depends on the trellis from $n-O$ to $n+O$ \cite{sliding}. With the overlap before and behind the data, the received symbols in different blocks can be processed in parallel. Therefore, the received symbols in different blocks are processed through parallel BM calculation, parallel ACS, and parallel survivor path selection. In our paper, all the MLSE algorithms described in the following are based on the parallel sliding block architecture. For the parallel MLSE described in this section, although the received symbols in different blocks are processed in parallel, the symbols within the same block are still processed one after another serially in ACS. To distinguish from the  processing method described in Section \ref{2-step}, the parallel MLSE described in this section is termed 1S-MLSE, as symbols within the same block are still processed one after another by 1-step.

For the architecture of 1S-MLSE, one of the parallel sliding blocks with the symbol length of $N=O+R+O$ is analyzed. The BM of all the $N$ received symbols can be calculated in parallel. The PM of the previous time must be transferred to the next time during the ACS operation (the initial PM for the first ACS is zero). Consequently, the prepared BM must wait sequentially. For the block with $N$ received symbols, the latency of ACS should be $N$ delay units, which increases with the increase of block length. Both the BM calculation and survivor path selection introduce the latency of $1$ delay unit, resulting in the overall latency of $N+2$ delay units for 1S-MLSE.

Finally, the computational complexity of 1S-MLSE with the symbol length of $N$ is studied. For the BM calculation, the channel response coefficients should be configurable. Considering that the channel response changes slowly, the equation $\alpha s_{n-1} +s_n$ can be calculated only once for every $N$ symbols. Therefore, the computational complexity of BM calculation includes $16N$ variable multipliers, 1 constant multiplier, and $16N+16$ adders. For the ACS calculation, $16(N-1)$ adders and $12N$ comparators are included. In the survivor path selection, the path with the minimal PM is chosen from the last 4 paths, which requires 3 comparators. In conclusion, the overall computational complexity of 1S-MLSE includes $16N$ variable multipliers, 1 constant multiplier, $32N$ adders, and $12N+3$ comparators.

\section{Low-Latency  MLSE} \label{2-step}
In the 1S-MLSE, symbols in the ACS calculation should be processed one after another. The latency of ACS for $N$ received symbols in a sliding block is $N$ delay units, which is increased with the increase of the value $N$. Inspired by the look-ahead $M$-step technique \cite{M1,M2,M3,M4,M5},  MLSE with parallel sliding block architecture can be applied to reduce the latency. The main difference between 1S-MLSE and  MLSE is in the process of ACS. In the 1S-MLSE, one group of $16$ BMs from time index $n-1$ to time index $n$ is processed by 4 ACS units at a time. In the  MLSE, two successive steps of $32$ BMs from time index $n-1$ to time index $n+1$ are processed by 16 ACS units in the first layer. After the first layer, the output of two groups of accumulated BM is still processed by 16 ACS units in the next layer. Finally, the last two groups of accumulated BM are input into 16 ACS units, and the survivor path is found from the outputs of these 16 ACS units. Owing to the parallel processing, it is obvious that the latency of ACS is reduced from $N$ delay units to $\log_2(N)$ delay units.

\textit{1) BM Calculation and ACS of the First Layer:} The signal processing procedure of  MLSE still includes BM calculation, ACS, and survivor path selection. The BM calculation of  MLSE is the same as that of 1S-MLSE. In the first layer, the BMs from time index $n-1$ to time index $n+1$ that share the same initial and final states are summed together. Since all four paths start from the same state, their BMs are comparable regardless of the previous PM. In each ACS unit, 4 adders and 3 comparators are included. In the overall 16 ACS units, 64 adders and 48 comparators are included. In the first layer with $N$ symbols, the overall groups of 16 ACS units are $N/2$. In conclusion, with symbol length of $N$, the overall computational complexity of the BM calculation and ACS operation in the first layer includes $16N$ variable multipliers, $1$ constant multipliers, $48N+16$ adders, and $24N$ comparators.

\textit{2) ACS of the Other Layers and Survivor Path Selection:} After the first layer, the accumulated BM are transferred layer by layer. The overall groups of 16 ACS units are decreased from the first layer of $N/2$ to the following layers of $N/4$, $N/8$, and so on. In the last layer, the survivor path is found from the outputs of the last group of 16 ACS units. The survivor path selection requires 15 comparators and introduces the latency of $1$ delay unit. Therefore, the overall latency of  MLSE is $\log_2(N)+2$ delay units. The overall computational complexity of  MLSE is $16N$ variable multipliers, 1 constant multiplier, $80N-48$ adders, and $48N-33$ comparators. Compared with 1S-MLSE, the latency of  MLSE is reduced from linear order to logarithmic order at the cost of increased adders and comparators.

\section{Simplified  MLSE} \label{2-step low complexity}

Since the latency of  MLSE is reduced at the cost of complexity, simplified  MLSE is proposed in this section. Simplified  MLSE consists of two parts: computational simplification and reduced state. Computational simplification is proposed to reduce the number of multipliers. Reduced state is combined with computational simplification, further reducing the number of adders and comparators.

\subsection{Computational Simplification}

\textit{1) BM Calculation and ACS of the First Layer:} To reduce the number of multipliers, the computational simplification is proposed, which combines the common terms in the BM calculation and addition operations of the first layer. The 64 expansion equations of the BM calculation and addition operations from time index $n-1$ to time index $n+1$ in the first layer are shown in Appendix A. These 64 expansion equations can be expressed as the sum of vectors ${\bf{A}}$, ${\bf{B}}$, ${\bf{C}}$, and ${\bf{D}}$. Both the vector ${\bf{A}} = [{A_1},{A_2},...,{A_{16}}]$ and the vector ${\bf{C}} = [{C_1},{C_2},...,{C_{16}}]$ are related to $\alpha$ and constant values. Both the vector ${\bf{B}} = [{B_1},{B_2},...,{B_{16}}]$ and the vector ${\bf{D}} = [{D_1},{D_2},...,{D_{16}}]$ are related to the received symbol $y$.

Through combining the common terms, the vectors ${\bf{A}}$ and ${\bf{C}}$ include 1 variable multiplier for $\alpha^2$, 4 constant multipliers for $10 \alpha^2$, $18 \alpha^2$, $12 \alpha$, $36 \alpha$, and additional 31 adders. Multiplication with powers of 2 can be achieved through bit-shift operations, thereby avoiding the use of multipliers. For the calculation of ${\bf{B}}$ and ${\bf{D}}$, ${\bf{B}}$ can be decomposed into ${\bf{M}}+{\bf{N}}+{\bf{J}}$, while ${\bf{D}}$ can be decomposed into ${\bf{M}}+{\bf{N}}+{\bf{K}}$. ${\bf{M}}$ corresponds to the common term $y_{n}+ \alpha y_{n+1}$ and its multiples. ${\bf{N}}$ corresponds to the common term $\alpha y_{n}$ and its multiples. ${\bf{J}}$ is equal to the common term $6y_{n+1}$ for ${\bf{B}}$, and ${\bf{K}}$ is equal to the common term $2y_{n+1}$ for ${\bf{D}}$.

Therefore, through combining the common terms, the vectors ${\bf{B}}$ and ${\bf{D}}$ include 2 variable multipliers, 3 constant multipliers, and 41 adders. The computation of ${\bf{B}}$ and ${\bf{D}}$ can be performed simultaneously with that of ${\bf{A}}$ and ${\bf{C}}$. In conclusion, with computational simplification, the overall complexity of BM calculation and ACS operation in the first layer includes $N+1$ variable multipliers, $3N/2+4$ constant multipliers, $105N/2+31$ adders, and $24N$ comparators. 

\textit{2) ACS of the Other Layers and Survivor Path Selection:} The difference between  MLSE with and without computational simplification only exists in the first layer, since the second layer can only utilize the computationally simplified results from the first layer. From the second layer to the last layer, the computing process of  MLSE with computational simplification is identical to that of  MLSE without computational simplification, which preserves the same latency and computational complexity.

To make a fair comparison, the computational simplification is also applied to 1S-MLSE. The expansion equations can be expressed as the sum of vectors ${\bf{E}}$ and ${\bf{F}}$. By combining common terms, the vector ${\bf{E}} = [{E_1},{E_2},...,{E_{8}}]$ includes 1 variable multiplier for $\alpha^2$, 3 constant multipliers for $9 \alpha^2$, $6 \alpha$, $18 \alpha$, and 12 adders. The vector ${\bf{F}} = [{F_1},{F_2},...,{F_{8}}]$ includes 1 variable multiplier for $\alpha y_{n}$, 2 constant multipliers for $6 \alpha y_{n}$ and $6y_{n}$, plus 8 adders. 16 adders are needed for the addition between ${\bf{E}}$ and ${\bf{F}}$, while 12 comparators are utilized after the addition operations. After the BM calculation with computational simplification, ACS and survivor path selection are the same as those of 1S-MLSE without computational simplification, which preserves the same latency and computational complexity.

In conclusion, the overall computational complexity of 1S-MLSE with computational simplification is $N+1$ variable multipliers, $2N+3$ constant multipliers, $40N-4$ adders, and $12N+3$ comparators, while that of  MLSE with computational simplification is $N+1$ variable multipliers, $3N/2+4$ constant multipliers, $169N/2-33$ adders, and $48N-33$ comparators. Both 1S-MLSE with computational simplification and  MLSE with computational simplification use the same number of variable multipliers. Because constant multipliers and comparators can be implemented by adders, variable multipliers occupy far more hardware resources than constant multipliers, adders, or comparators. When using computational simplification, the number of variable multipliers is reduced from $O(NM^{L})$ to $O(N)$. The BER performance of 1S-MLSE, 1S-MLSE with computational simplification,  MLSE, and  MLSE with computational simplification is analyzed. All four algorithms have the same BER performance. By combining common items, the calculation accuracy remains unchanged, ensuring the computational simplification does not degrade the BER performance.

\subsection{Simplified  MLSE with the Combination of Computational Simplification and Reduced State} \label{simplified}

To reduce the number of adders and comparators, the reduced state is combined with computational simplification. In our paper, we refer to  MLSE with the combination of computational simplification and reduced state as simplified  MLSE. The output signal $d$ of FFE can be regarded as the pre-decision value of the received signal \cite{hoon}. For PAM4 with 4 states of $\{-3,-1,1,3\}$, if the number of states is reduced from 4 to 3 through pre-decision value, the possible states include $\{-3,-1,1\}$ or $\{-1,1,3\}$, which are represented by $\{a,b,c\}$. If the number of states is reduced from 4 to 2 through pre-decision value, the possible states include $\{-3,-1\}$, $\{-1,1\}$ or $\{1,3\}$, which are represented by $\{a,b\}$. When the state number is reduced to 3, the number of ACS units in a  calculation is reduced from 16 to 9, in which only 27 adders and 18 comparators are needed. When the state number is reduced to 2, the number of ACS units in a  calculation is reduced from 16 to 4, in which only 8 adders and 4 comparators are needed.

\textit{1) BM Calculation and ACS of the First Layer:} For the first layer, although the reduced state is applied, all the possible state sets can be encountered. Therefore, all the equations in ${\bf{A}}$, ${\bf{B}}$, ${\bf{C}}$ and ${\bf{D}}$ should be calculated in advance. There is no difference for the calculation of ${\bf{A}}$, ${\bf{B}}$, ${\bf{C}}$ and ${\bf{D}}$ no matter whether the reduced state is applied. When calculating ${\bf{A}}+{\bf{B}}$, ${\bf{A}}-{\bf{B}}$, ${\bf{C}}+{\bf{D}}$ or ${\bf{C}}-{\bf{D}}$, the number of adders is reduced to 27 for 3 states and 8 for 2 states. After the adders, the number of comparators is reduced to 18 for 3 states and 4 for 2 states. In conclusion, for 3 states, the computational complexity of the first layer includes $N+1$ variable multipliers, $3N/2+4$ constant multipliers, $34N+31$ adders, and $9N$ comparators. For 2 states, the computational complexity of the first layer includes $N+1$ variable multipliers, $3N/2+4$ constant multipliers, $49N/2+31$ adders, and $2N$ comparators.

\textit{2) ACS of the Other Layers and Survivor Path Selection:} From the second layer to the last layer, due to the decrease in ACS units, 27 adders and 18 comparators are needed for a  calculation with 3 states. 8 adders and 4 comparators are needed for a  calculation with 2 states. In the final survivor path selection step, the path with the minimal accumulated metrics is chosen from 9 paths for the state number of 3, reducing the number of comparators to 8. Similarly, the path with the minimal accumulated metrics is chosen from 4 paths for the state number of 2, reducing the number of comparators to 3.

We compares the BER performance of  MLSE under three configurations: all 4 states (with only computational simplification), simplified  MLSE with 3 states, and simplified  MLSE with 2 states. In our experiment, the noise enhancement caused by FFE is not severe enough to damage the pre-decision value from FFE for 3 or 2 reduced states. Therefore, the BER performance remains nearly identical regardless of whether all 4 states, 3 states, or 2 states are used. For the system with a larger $\alpha$ value, due to the larger noise in the pre-decision value, the BER performance of 2 states may have worse BER performance than 3 states \cite{hoon}. Owing to the nearly identical BER performance, the simplified scheme in our paper refers to the combination of computational simplification and 2 reduced states. To make a fair comparison, the simplified scheme is also applied to 1S-MLSE.

\section{Experimental Setup} \label{setup}
The experiment of a 112-Gbit/s PAM4 signal over 2-km SSMF transmission in C-band is performed to verify the performance of the proposed MLSE. At the transmitter, $4 \times 10^5$ pseudo-random binary sequences (PRBS) are mapped to $2 \times 10^5$ PAM4 symbols, in which the first 1000 symbols are served as training symbols. To improve the transmission performance, the optimized pre-equalization method proposed in our previous work is applied to compensate for partial bandwidth-limited distortion \cite{guo}. The PAM4 signal is pulse-shaped by a root-raised cosine (RRC) filter with the roll-off factor of 0.1. Then, the signal is resampled by 1.25 sps to align with the sampling rate of the digital-to-analog converter (DAC). The 56-GBaud PAM4 analog signal is generated by a DAC with a sampling rate of 70 GSa/s. The 3-dB bandwidth of the 8-bit DAC is 16 GHz. Afterwards, the analog signal is amplified by a 40-Gbps electrical amplifier (EA, CENTELLAX OA4MVM3). A 40-Gbps optical electro-absorption modulator (EAM, OM5757C-CTM388) is employed to modulate the electrical signal on an optical carrier at $\sim$1550 nm. The DC bias applied to the EAM is $-$1.6 V, and the insertion loss of the EAM is $\sim$9 dB.

After 2-km SSMF transmission, the received optical power (ROP) is adjusted by a variable optical attenuator (VOA). Then, the optical signal is converted to an electrical signal through a 30-Gbps PIN photodiode with an integrated trans-impedance amplifier (PIN-TIA). Finally, the electrical signal is converted to a digital signal by a 70 GSa/s analog-to-digital converter (ADC). The offline digital signal processing includes matched filter, resampling, synchronization, FFE, post-filter, MLSE, PAM4 demapping, and BER calculation. Under 2-km SSMF transmission, the 10-dB bandwidth of the whole system is only $\sim$15 GHz. Meanwhile, the frequency response drops rapidly beyond 21 GHz, which leads to severe bandwidth-limited distortion on the PAM4 signal.

\section{Experimental Results and Discussions} \label{result}
\subsection{Parameter Configuration for MLSE}
The inputs of MLSE include the coefficients of the post-filter and the signal after the post-filter. For the 2-tap post-filter with the coefficients of $[1, \alpha]$, the optimal value of $\alpha$ is swept to obtain the best BER performance after MLSE. The conventional 1S-MLSE is applied to determine the value of $\alpha$, because the conventional 1S-MLSE serves as a performance benchmark for the other proposed MLSE algorithms. After 2-km SSMF transmission, to achieve the best BER performance, the optimal $\alpha$ value is set to 0.55. This value is not too large owing to the use of pre-equalization.

After determining the value of $\alpha$, the number of received symbols $R$ and overlapped symbols $O$ should be determined for each sliding block. For serial MLSE, the PM of the previous block is transferred to the next block. The optimal performance is achieved through the transfer of PM, so that each block only includes received symbols of length $R$ and post-overlap of length $O$. For the parallel MLSE used in our paper, the PM is not transferred to the following blocks. Each block includes pre-overlap of length $O$, received symbols of length $R$, and post-overlap of length $O$. Because the parallel MLSE uses the accumulated BM of pre-overlap to serve as the PM in serial MLSE, the principle to determine the value of $R$ and $O$ is that the parallel MLSE has almost the same BER performance as the conventional serial MLSE. The 1S-MLSE is still applied to determine the value of $R$ and $O$.

In our experiment, the number of received symbols $R$ is set to 16, since smaller values impair the performance of parallel MLSE. When the number of overlapped symbols is equal to or larger than 8, parallel MLSE can achieve almost the same BER performance as serial MLSE. When the number of overlapped symbols is less than 8, parallel MLSE has worse BER performance than serial MLSE. Therefore, in the parallel MLSE architecture of our paper, the number of symbols $N$ in a sliding block is set to 32, which includes 8 symbols of pre-overlap, 16 symbols of useful data, and 8 symbols of post-overlap. All the algorithms described in our paper are applied with this parallel architecture.

\subsection{BER Performance, Latency and Complexity} 

After determining the parallel architecture, the experimental results of 112-Gbit/s PAM4 after 2-km SSMF transmission are analyzed. As described in Section \ref{simplified}, the term simplified  MLSE in our paper refers to  MLSE with the combination of computational simplification and 2 reduced states. At the ROP of $-$7.5 dBm, the BER performance of 112-Gbit/s PAM4 after 2-km SSMF transmission achieves the 7\% FEC limit with simplified  MLSE, whereas it fails to meet the 7\% FEC limit when only FFE is applied. If the ROP is further increased, the BER performance becomes worse due to the saturation of the TIA. With the symbol length $N=32$, the latency and complexity of 1S-MLSE, simplified 1S-MLSE,  MLSE, and simplified  MLSE are analyzed.  MLSE architecture can reduce the latency from 34 delay units in linear order to 7 delay units in logarithmic order.

When considering computational complexity, both constant multipliers and comparators can be implemented using adders. Thus, variable multipliers occupy far more hardware resources than constant multipliers, adders, or comparators. In simplified 1S-MLSE and simplified  MLSE, the number of variable multipliers is reduced from 512 in exponential order to 33 in linear order. Compared with  MLSE, the number of adders and comparators for simplified  MLSE is reduced to 37.2\% and 8.4\%, respectively. The number of adders and comparators in  MLSE is 2.45 times and 3.88 times that of 1S-MLSE, respectively, while the number of adders and comparators in simplified  MLSE drops to 1.80 times and 1.95 times that of simplified 1S-MLSE, respectively. In conclusion, the simplified scheme can reduce the utilization of multipliers, adders, and comparators.

\section{Conclusion} \label{conclusion}
In this paper, we propose and experimentally verify the simplified  MLSE algorithm. The simplified scheme combines computational simplification and reduced state. Computational simplification does not degrade the BER performance. The noise enhancement caused by FFE is not severe enough to damage the pre-decision value in our experiment. Therefore, the use of reduced state maintains nearly identical performance. In the experiment of 112-Gbit/s PAM4 transmission under 2-km SSMF, the simplified  MLSE can achieve better BER performance than FFE, enabling the BER performance to reach the 7\% FEC limit. Compared with simplified 1S-MLSE, the latency of simplified  MLSE is decreased from 34 delay units in linear order to 7 delay units in logarithmic order. Compared with  MLSE, the simplified  MLSE reduces the number of variable multipliers from 512 in exponential order to 33 in linear order, while the number of adders and comparators is reduced to 37.2\% and 8.4\%, respectively. In conclusion, the proposed simplified  MLSE shows great potential for future high-speed, low-latency, and low-complexity optical interconnects.

\ifCLASSOPTIONcaptionsoff
  \newpage
\fi



%
\bibliography{reference}

\end{document}